\documentstyle[multicol,epsf,aps]{revtex}
\begin{document}
\title{Stationary Velocity Distributions in Traffic Flows}
\author{E.~Ben-Naim$\dag$ and P.~L.~Krapivsky$\ddag$}
\address{$\dag$Theoretical Division and Center for Nonlinear Studies, 
Los Alamos National Laboratory, Los Alamos, NM 87545}
\address{$\ddag$Center for Polymer Studies and Department of Physics,
Boston University, Boston, MA 02215}
\maketitle
\begin{abstract} 
  
  We introduce a traffic flow model that incorporates clustering and
  passing. We obtain analytically the steady state characteristics of
  the flow from a Boltzmann-like equation.  A single dimensionless
  parameter, $R=c_0v_0t_0$ with $c_0$ the concentration, $v_0$ the
  velocity range, and $t_0^{-1}$ the passing rate, determines the
  nature of the steady state. When $R\ll 1$, uninterrupted flow with
  single cars occurs. When $R\gg 1$, large clusters with average mass
  $\langle m\rangle\sim R^{\alpha}$ form, and the flux is 
  $J\sim R^{-\gamma}$.  The initial distribution of slow cars governs
  the statistics. When $P_0(v)\sim v^{\mu}$ as $v\to 0$, the scaling
  exponents are $\gamma=1/(\mu+2)$, $\alpha=1/2$ when $\mu>0$, and
  $\alpha=(\mu+1)/(\mu+2)$ when $\mu<0$.


\end{abstract}

\begin{multicols}{2} 
\section{Introduction}
  
Traffic flows are strongly interacting many-body systems.  They also
present a natural testbed for theories and techniques developed for
physical systems such as kinetic theory and hydrodynamics.  Traffic
systems has been receiving much attention recently \cite{Gartner}, and
a number of approaches were suggested including fluid mechanics
\cite{Prigogine,kerner}, cellular automata
\cite{Nagel,Biham,Nagatani,Schadschneider,Nagel1,Nagel2,Brankov,Ktitarev}, 
particle hopping \cite{Der,Benjamini,Evans,Popkov}, and ballistic motion 
\cite{eps,Helbing,Nagatani1}.
Traffic networks can be viewed as low dimensional systems. For
example, rural traffic is intrinsically one-dimensional and urban grid
traffic is two-dimensional. Despite this important simplifying
feature, most studies in this area are numeric in nature.

Ballistic models are harder to simulate than cellular automata and
particle hopping models. However, they are more realistic since time and
space are treated as continuous variables.  They can also prove useful
for analytical treatment.  An exactly solvable clustering process
shows that extremal properties of the velocity distribution determine
the kinetic behavior \cite{eps}.  However, it results in ever-growing
and ever-slowing jams with a trivial steady state in a finite system.
In this study, we investigate more realistic situations where fast
cars can pass slow cars. This is motivated by and should be applicable
to passing zones of one lane roadways as well as multilane highways.
Our goal is to determine analytically statistical properties of the
flow such as the flux, and characterize their dependence on the
intrinsic velocity distribution.

\begin{figure}
\vspace{-.1in}
\centerline{\epsfxsize=7cm \epsfbox{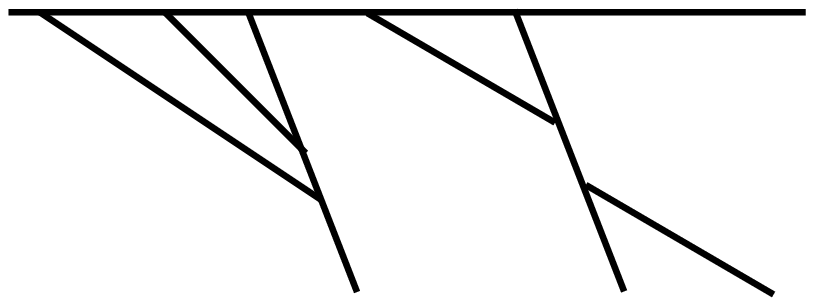}}
\noindent{\small {\bf Fig.1}
Space time diagram of the traffic model. 
Formation of a cluster with two fast cars is shown to the left
and formation of a one car cluster and its breakup due to
escape is shown to the right.}
\end{figure}

We start by formulating the model. Consider a one-dimensional traffic
flow with sizeless cars (``particles'') moving with a constant
velocity. We assume that cars have intrinsic velocities by which
they would drive on an empty road.  Initially, cars are randomly
distributed in space and they drive with their intrinsic velocities.
However, the presence of slower cars forces some cars to drive behind
a slower car and therefore leads to the formation of clusters.  Simple
collision and escape mechanisms are implemented. When a cluster
overtakes a slower cluster, a larger cluster forms. It moves with the
smaller of the two velocities.  Meanwhile, all cars in a given cluster
may escape their respective cluster and resume driving with their
intrinsic velocity.  We assume a constant escape rate $t_0^{-1}$.  The
actual collision and escape times are proportional to the car size and
thus set to zero (these time scales should become important in heavy
traffic).
  
A heuristic argument suggests that a single dimensionless parameter
underlies the steady state. Consider a state where the car
concentration is $c_0$, and the typical intrinsic velocity range is
$v_0$.  Let the steady state cluster density be $c<c_0$, which implies
the typical cluster size $\langle m\rangle=c_0/c$. If large clusters
form, $\langle m\rangle\gg 1$, then the overall escape rate can be
estimated by $\langle m\rangle t_0^{-1}$. Assuming that most collisions
involve fast cars and slow clusters, the typical collision rate is
$cv_0$. In the steady state, the number of cars joining and leaving
clusters should balance and thus, $c_0/(ct_0)=v_0 c$ or
$c=(c_0/v_0t_0)^{1/2}$.  This heuristic argument gives the leading
behavior of the average cluster size

\begin{equation}
\label{m}
\langle m\rangle\sim R^{1/2}\qquad {\rm when}\quad R\gg 1, 
\end{equation}
where $R$ is the ratio of the two elementary time scales, the escape
time $t_{\rm esc}=t_0$ and the collision time $t_{\rm col}=(c_0v_0)^{-1}$:

\begin{equation}
\label{R}
R={t_{\rm esc}\over t_{\rm col}}=c_0v_0t_0. 
\end{equation}
We term this dimensionless quantity the ``collision number''.  For
large collision numbers, large clusters occur according to
Eq.~(\ref{m}), while for small collision numbers the effect of
collisions is small $\langle m\rangle\cong 1+aR$.  Analysis of the
master equations detailed below confirms this heuristic picture
under quite general conditions. 

The rest of this paper is organized as follows. In Sec.~II, the
master equations are used to derive analytical
expressions for various velocity distributions in the steady state. The
leading behavior in the limiting cases of light and heavy traffic are
highlighted in Sec.~III. Explicit expressions are written for the
special cases of uniform initial and final velocity distributions as
well as discrete distributions in Sec.~IV. We close with some open
problems, a discussion, and possible  applications.

\section{Theory}

In the following, it is convenient to introduce dimensionless velocity
$v/v_0\to v$, space $xc_0\to x$, and time $c_0v_0t \to t$ variables.
This rescales the escape rate $t_0^{-1}$ to the inverse collision
number $R^{-1}$. Let $P(v,t)$ be the density of clusters moving with
velocity $v$ at time $t$.  Initially, isolated single cars drive with
their intrinsic velocities drawn from the distribution $P_0(v)\equiv
P(v,t=0)$. This intrinsic velocity distribution is normalized to
unity, $\int dv P_0(v)=1$. The flow is invariant under a velocity
translation, and the minimal velocity is set to zero.

Initially, the velocities and the positions of the particles are
uncorrelated. Escape effectively mixes the positions and the
velocities.  Assuming that no spatial correlations develop, a closed
master equation for the velocity distribution of clusters $P(v,t)$ can
be written
\begin{eqnarray}
\label{pvt}
{\partial P(v,t)\over \partial t}=&&R^{-1}\left[P_0(v)-P(v,t)\right]\\
-&&P(v,t)\int_0^v dv' (v-v')P(v',t).\nonumber
\end{eqnarray}
The density of slowed down cars with intrinsic velocity $v$ is
$P_0(v)-P(v,t)$. Such cars escape their clusters with rate $R^{-1}$, and
thus the escape term.  Collisions occur with rate proportional to the
velocity difference as well as the product of the velocity
distributions.  The integration limits ensure that only collisions with
slower cars are taken into account.  

Steady state is obtained by taking the long time limit $t\to\infty$ or
$\partial/\partial t=0$.  Since we are primarily interested in the
steady state, we omit the time variable $P(v)\equiv
P(v,t=\infty)$.  Equating the right-hand side of the master equation to
zero, a relation between the intrinsic car distribution and steady state
cluster distribution emerges

\begin{equation}
\label{pv}
P(v)\left[1+R\int_0^v dv'(v-v')P(v')\right]=P_0(v).
\end{equation}
Given the intrinsic velocity distribution this relation gives the
final cluster velocity distribution only implicitly. In contrast, the
inverse problem is simpler as knowledge of the final distribution, the
observed quantity in real traffic flows, gives explicitly the
intrinsic distribution.  We confirm that in the limit $R\to\infty$,
all clusters move with the minimal velocity $P(v)\to \delta(v)$, while
in the limit $R\to 0$, all cars move with their intrinsic velocity
$P(v)\to P_0(v)$. 

It is convenient to transform the integral equation (\ref{pv}) into a 
differential one.  Consider the  auxiliary function
\begin{equation}
\label{Q}
Q(v)=R^{-1}+\int_0^v dv'(v-v')P(v'),
\end{equation}
which gives the cluster distribution by second differentiation 
\begin{equation}
\label{PQ}
P(v)=Q''(v). 
\end{equation}
Thence, the steady state condition (\ref{pv}) reduces to the second
order nonlinear differential equation 
\begin{equation}
\label{Qrho}
Q(v)Q''(v)=R^{-1}P_0(v). 
\end{equation}
The boundary conditions are $Q(0)=R^{-1}$ and $Q'(0)=0$. 
The cluster concentration is found from the cluster velocity
distribution using 
\begin{equation}
\label{cdef}
c=\int_0^{\infty}  dv\, P(v),
\end{equation}
and the average cluster mass is simply $\langle m\rangle=c^{-1}$.
Furthermore, the average cluster velocity is obtained from 
\begin{equation}
\label{vavdef}
\langle v\rangle=c^{-1}\int_0^{\infty} dv \, v P(v).
\end{equation}

Cars may drive with a velocity smaller than their intrinsic one, and
it is natural to consider the joint velocity distribution $P(v,v')$,
the density of cars of intrinsic velocity $v$ driving with velocity
$v'$.  The master equation for the joint distribution reads

\begin{eqnarray}
\label{pvvt}
{\partial P(v,v')\over \partial t}
=&-&R^{-1}P(v,v')+(v-v')P(v)P(v')\nonumber\\
&-&P(v,v')\int_0^{v'} dv'' (v'-v'')P(v'')\\
&+&P(v')\int_{v'}^v dv''(v''-v')P(v,v'').\nonumber
\end{eqnarray}
The first term accounts for loss due to escape, while the rest of the
terms represent changes due to collisions. For instance, the last term
describes events where a $v$-car driving with velocity $v''$ is
further slowed down after a collision with a $v'$-cluster.  One can
verify that the total number of $v$-cars, 
\begin{equation}
\label{norm}
P_0(v)=P(v)+\int_0^v dv' P(v,v'),
\end{equation}
is conserved by the evolution Eqs.~(\ref{pvt}) and (\ref{pvvt}).

At the steady state, the joint distribution satisfies 
\begin{equation}
\label{QP}
P(v,v')Q(v')=(v-v')P(v)P(v')+Q(v,v')P(v'), 
\end{equation}
obtained using the definition of $Q(v)$ and the joint auxiliary function 
\begin{equation}
\label{Qvv}
Q(v,v')=\int_{v'}^v dw(w-v')P(v,w).
\end{equation}
Although the collision number $R$ does not appear in Eq.~(\ref{QP})
explicitly, it enters through $Q(v)$ and $P(v)$.  

Combining (\ref{QP}) with Eqs.~(\ref{Qvv}), (\ref{PQ}), and using the
relationship $P(v,v')=\partial^2 Q(v,v')/\partial {v'}^2$ yields 
\begin{equation}
\label{QP1}
{\partial \over \partial v'}
\left[Q^2(v')\,{\partial \over \partial v'}\,
{Q(v,v')\over Q(v')}\right]=(v-v')P(v)P(v').
\end{equation}
Integrating twice over $v'$ gives the joint auxiliary function in
terms of the single variable functions 
\begin{equation}
\label{QP3'}
Q(v,v')=P(v)Q(v')\int_{v'}^v\!\!{du\over
  Q^2(u)}\int_u^v\!\!dw(v-w)P(w). 
\end{equation}
The boundary conditions $Q(v,v)={\partial \over \partial
  v'}Q(v,v')\big|_{v'=v} =0$ were used to obtain this expression. 
Furthermore, integration by parts of $\int_u^v dw (v-w)P(w)=\int_u^v
dw (v-w)Q''(w)$ gives

\begin{equation}
\label{QP3}
Q(v,v')=P(v)\left[Q(v)Q(v')\int_{v'}^v {du\over Q^2(u)}-(v-v')\right].
\end{equation}
Substituting Eq.~(\ref{QP3}) into (\ref{QP}) and then replacing $PQ$
with $R^{-1}P_0$ we find a relatively simple expression for the joint
velocity distribution
\begin{equation}
\label{pvv}
P(v,v')={P_0(v)P_0(v')\over Q(v')}\int_{v'}^v {du\over [RQ(u)]^2}. 
\end{equation}

Another interesting quantity is the flux or the average velocity given
by \hbox{$J=\int dv \left[vP(v)+\int_0^v dw\, w P(v,w)\right]$}. From
the definition of the joint auxiliary function, the second integral is
identified with $Q(v,0)$, implying 
\begin{equation}
\label{jdef}
J=\int_0^{\infty} dv\left[vP(v)+Q(v,0)\right].  
\end{equation}
The integrand can be considerably simplified using Eq.~(\ref{QP3}),
$Q(0)=R^{-1}$, and Eq.~(\ref{Qrho}). The term $vP(v)$ cancels and we
find a useful expression for the flux 

\begin{equation}
\label{j}
J=\int_0^\infty dv\,P_0(v) \int_0^v {du\over [RQ(u)]^2}.
\end{equation}

One can also ask for the actual velocity distribution of cars defined via
\begin{equation}
\label{gdef}
G(v)=P(v)+\int_v^{\infty}\!\! dw\,P(w,v). 
\end{equation}
Substituting the
joint velocity distribution allows us to express the car velocity
distribution via single variable distributions 

\begin{equation}
\label{g}
G(v)=P(v)\left[1+R\int_v^\infty dw\,P_0(w)\int_v^w {du\over [RQ(u)]^2}\right].
\end{equation}
The car velocity distribution satisfies the normalization conditions
\hbox{$1=\int\!dv\, G(v)$} and \hbox{$J=\int\! dv \, v G(v)$}.

In summary, for arbitrary intrinsic velocity distributions, the entire
steady state problem is reduced to the nonlinear second order
differential equation (\ref{Qrho}). Given $Q(v)$, steady state
characteristics such as $P(v)$, $P(v,v')$, $J$, and $G(v)$ can be
calculated using the {\rm explicit} formulae (\ref{PQ}), (\ref{pvv}),
(\ref{j}), and (\ref{g}), respectively.

\section{Limiting cases}

Although one cannot solve Eq.~(\ref{Qrho}) analytically in general, it
is still possible to obtain the leading behavior in the limits of 
$R\to 0$ and $R\to \infty$.

\subsection{Low Collision Numbers}

To analyze the flow characteristics in the collision-controlled
regime, $R\ll 1$, we use Eq.~(\ref{pv}) to write $P(v)$ as a perturbation
expansion in $R$:
\begin{equation}
\label{pv1}
P(v)\cong P_0(v)\left[1-R\int_0^v\!\!\!dv'(v-v')P_0(v')\right].
\end{equation}
In this limit, the auxiliary function is roughly constant $RQ(v)\cong 
1$,  and Eq.~(\ref{pvv}) gives the joint distribution to first
order in $R$ 

\begin{equation}
\label{pvv1}
P(v,v') \cong R(v-v')P_0(v)P_0(v').
\end{equation}
The final density and flux are

\begin{equation}
\label{cj1}
c\cong 1-c_1 R, \quad
J\cong J_0-J_1 R,
\end{equation}
with $c_1=\int dv P_0(v)\int_0^v dv'(v-v')P_0(v')$, $J_0=M_1$,
$J_1=M_2-M_1^2$ ($M_n$ are the moments of the intrinsic velocity
distribution $M_n=\int dv\,v^n P_0(v)$).  The coefficient $J_1\geq 0$
equals the width of the initial velocity distribution. This gives a
simple intuitive picture: the larger the initial velocity
fluctuations, the smaller the flux.  By either substituting the joint velocity
distribution into the definition of $G(v)$, or from 
Eq.~(\ref{g}), the car velocity distribution is

\begin{equation}
\label{g1}
G(v)\cong P_0(v)\left[1+R\int_0^\infty dv' (v'-v)P_0(v')\right]. 
\end{equation}
As the integral is over the entire velocity range, the order $R$
correction is positive for small $v$ and negative for large $v$. In
other words $G(v)>P_0(v)$ when $v<v_c$. The crossover velocity equals
the average intrinsic velocity $v_c=J_0=M_1$, as seen from
Eq.~(\ref{g1}). 

We conclude that the collision-controlled limit is weakly interacting,
explicit expressions for the leading corrections of the steady state
properties are possible.

\subsection{Large Collision Numbers}

The analysis in the complementary escape-controlled regime, $R\gg 1$,
is more subtle since the condition $R\int_0^v dv'
(v-v')P_0(v')\ll 1$ is satisfied only for small
velocities.  No matter how large $R$ is, sufficiently slow cars
are not affected by collisions, and $P(v)$ is given by
Eq.~(\ref{pv1}) when $v\ll v^*$. The threshold velocity $v^*\equiv
v^*(R)$ is estimated  from \hbox{$R\int_0^{v^*} dv(v^*-v)P_0(v)\sim 1$}. 

It is useful to consider algebraic intrinsic distributions

\begin{equation}
\label{mu}
P_0(v)=(\mu+1)v^{\mu} \quad \mu>-1,
\end{equation}
in the velocity range [0:1] with the prefactor ensuring unit
normalization.  For such distributions, the threshold velocity
decreases with growing $R$ according to $v^* \sim R^{-{1\over
    \mu+2}}$.  For $v\gg v^*$, the integral in Eq.~(\ref{pv})
dominates over the constant factor and \hbox{$R P(v)\int_0^v dv'
  (v-v') P(v')\sim v^{\mu}$}. Anticipating an algebraic behavior for
the cluster velocity distribution, $P(v)\sim R^{\sigma}v^{\delta}$
when $v\gg v^*$, gives different answers for positive and negative
$\mu$.  The leading behavior for $v\gg v^*$ can be summarized as
follows

\begin{equation}
\label{pvlead}
P(v)\sim\cases{R^{-1/(\mu+2)}v^{\mu-1}&$\mu<0$;\cr 
               R^{-{1\over 2}}v^{-1}[\ln(v/v^*)]^{-{1\over 2}}&$\mu=0$;\cr
               R^{-{1\over 2}}v^{{\mu\over 2}-1}&$\mu>0$.}
\end{equation}
The small and large velocity components of $P(v)$ match at the
threshold velocity, $P(v^*)\sim P_0(v^*)$.  Careful analysis, detailed
in the following section, is needed to get the logarithmic corrections
in the borderline case $\mu=0$. Substituting the leading asymptotic
behavior of Eq.~(\ref{pvlead}) into Eq.~(\ref{cdef}), the average
cluster size is found

\begin{equation}
\label{mav}
\langle m\rangle\sim\cases{R^{(\mu+1)/(\mu+2)} &$\mu<0$;\cr
                           (R/\ln R)^{1/2}     &$\mu=0$;\cr
                           R^{1/2}             &$\mu> 0$.}
\end{equation}
Similarly, the average cluster velocity defined in Eq.~(\ref{vavdef}) 
is evaluated 
\begin{equation}
\label{vav}
\langle v\rangle\sim\cases{R^{\mu/(\mu+2)}     &$\mu<0$;\cr
                           1/\ln R             &$\mu=0$;\cr
                           {\rm const}         &$\mu> 0$.}
\end{equation}

Two distinct regimes of behavior emerge.  For $\mu>0$, car-cluster
collisions dominate while for $\mu<0$ cluster-cluster collisions
dominate.  The scaling argument given in the introduction assumes the
former picture, and thus it does not hold in general. {\it A
  posteriori}, one can extend the scaling argument to the $\mu<0$
regime. The argument becomes involved, and we do not present it here.
Interestingly, in the cluster-cluster dominated regime, the scaling
behavior for the average cluster size, $\langle m\rangle\sim
R^{\alpha}$ with $\alpha=(\mu+1)/(\mu+2)$, is identical to the {\em
  kinetic} scaling, $\langle m\rangle\sim (c_0v_0t)^{\alpha}$ with the
same $\alpha$, found in the no passing limit \cite{eps}.  This
suggests an analogy between the dimensionless collision number
$R=c_0v_0t_0$ and the dimensionless time $c_0v_0t$.  The flux can be
evaluated in a similar fashion using Eq.~(\ref{j}),
\begin{equation}
\label{j2}
J\sim v^* \sim R^{-{1\over \mu+2}}.
\end{equation}
Interestingly, the flux is proportional to the threshold velocity
$v^*$. As a result, the flux exponent $\gamma=1/(\mu+2)$ is a regular
function of $\mu$ unlike the cluster size exponent $\alpha$.
Eq.~(\ref{j2}) is also consistent with identification of the crossover
velocity $v_c$ with the marginal velocity $v^*$.  No flux reduction
occurs when the intrinsic distribution is dominated by fast cars,
i.e., in the limit $\mu\to\infty$. In the other extreme, the maximal
flux reduction $J\sim R^{-1}$ is realized when $\mu\to -1$.

The car velocity distribution is strongly enhanced in the low
velocity limit, as seen by evaluating Eq.~(\ref{g})
\begin{equation}
\label{g2}
G(v)\sim R^{\mu+1\over \mu+2}v^{\mu}(1-{\rm const.}\times
v^{\mu+1}),\quad v\ll v^*.
\end{equation}
As a check of self-consistency, one can verify that \hbox{$1\sim
  \int_0^{v^*} dv\, G(v)$}, and \hbox{$J\sim v^* \sim \int_0^{v^*}
  dv\, v\, G(v)$}. Near the maximal velocity, the car velocity distribution
approaches the cluster distribution $G(v)\cong P(v)$.

In summary, as $R\to\infty$ the solution to the differential equation
(\ref{Qrho}) exhibits a boundary layer structure.  Inside the boundary
layer, $v<v^*$, the cluster velocity distribution is only slightly affected
by collisions, while in the outer region $v>v^*$,
the cluster velocity distribution is much smaller than the 
intrinsic velocity distribution.  The threshold velocity $v^*$
is determined by the small velocity behavior of the intrinsic velocity
distribution, and for the algebraic distributions (\ref{mu}) we have
found $v^*\sim R^{-{1\over \mu+2}}\to 0$.  The behavior detailed above in
the escape controlled limit is not restricted to  purely algebraic
distributions but is quite general.  We conclude that
a single parameter

\begin{equation}
\label{mudef}
\mu=\lim_{v\to 0} v{\partial\over\partial v } \ln P_0(v)
\end{equation}
determines the behavior as $R\to\infty$. In short, extreme statistics
underly the escape-limited flow properties.  Additionally, an
interesting transition between a slow and a fast velocity dominated
flow occurs at $\mu=0$.

\section{Examples}

Although the above analysis is quite general, it applies only to the
limiting values of $R$. To examine intermediate behavior, it is also
useful to obtain explicit solutions for some special cases.  Below, we
consider two relevant cases: uniform $P_0(v)$ and $P(v)$.  We also
obtain explicit expressions in the case of discrete velocity
distributions.

\subsection{Uniform Intrinsic Distribution}

We now consider the case of a uniform intrinsic distribution,
$P_0(v)=1$ for $0<v<1$. This case appears to be the most relevant to
real traffic flows since the intrinsic velocity distribution should
be regular near the the minimal velocity.  Integrating $QQ''=R^{-1}$
subject to the boundary conditions $Q(0)=R^{-1}$ and $Q'(0)=0$ gives
$Q'=\sqrt{2R^{-1}\ln (RQ)}$. Second integration gives
\begin{equation}
\label{q}
\int_1^{RQ}{dq\over \sqrt{2\ln q}}=v\sqrt{R},
\end{equation}
and thus implicitly determines $Q(v)$.  Evaluating the leading
behavior when $R\gg 1$, we find 
\begin{equation}
\label{asympt}
\langle m\rangle \simeq \sqrt{R\over \ln R}, \quad
\langle v\rangle \simeq {1\over \ln R}, \quad
J\simeq \sqrt{\pi\over 2R}.
\end{equation}
Fig.~2 shows the velocity distribution obtained numerically using
Eqs.~(\ref{q}) and (\ref{g}) for $R=10$.  For $v\ll v^*$, $G(v)\gg
P_0(v)$, and for $v\gg v^*$, \hbox{$G(v)\cong P(v)\ll P_0(v)$}. The
calculated distributions are consistent with the predictions, $G(0)\sim
R^{1/2}$ and $v^*\sim R^{-1/2}$. The car velocity distribution is 
linear near the origin in agreement with Eq.~(\ref{g2}).

\begin{figure}
\centerline{\epsfxsize=9cm \epsfbox{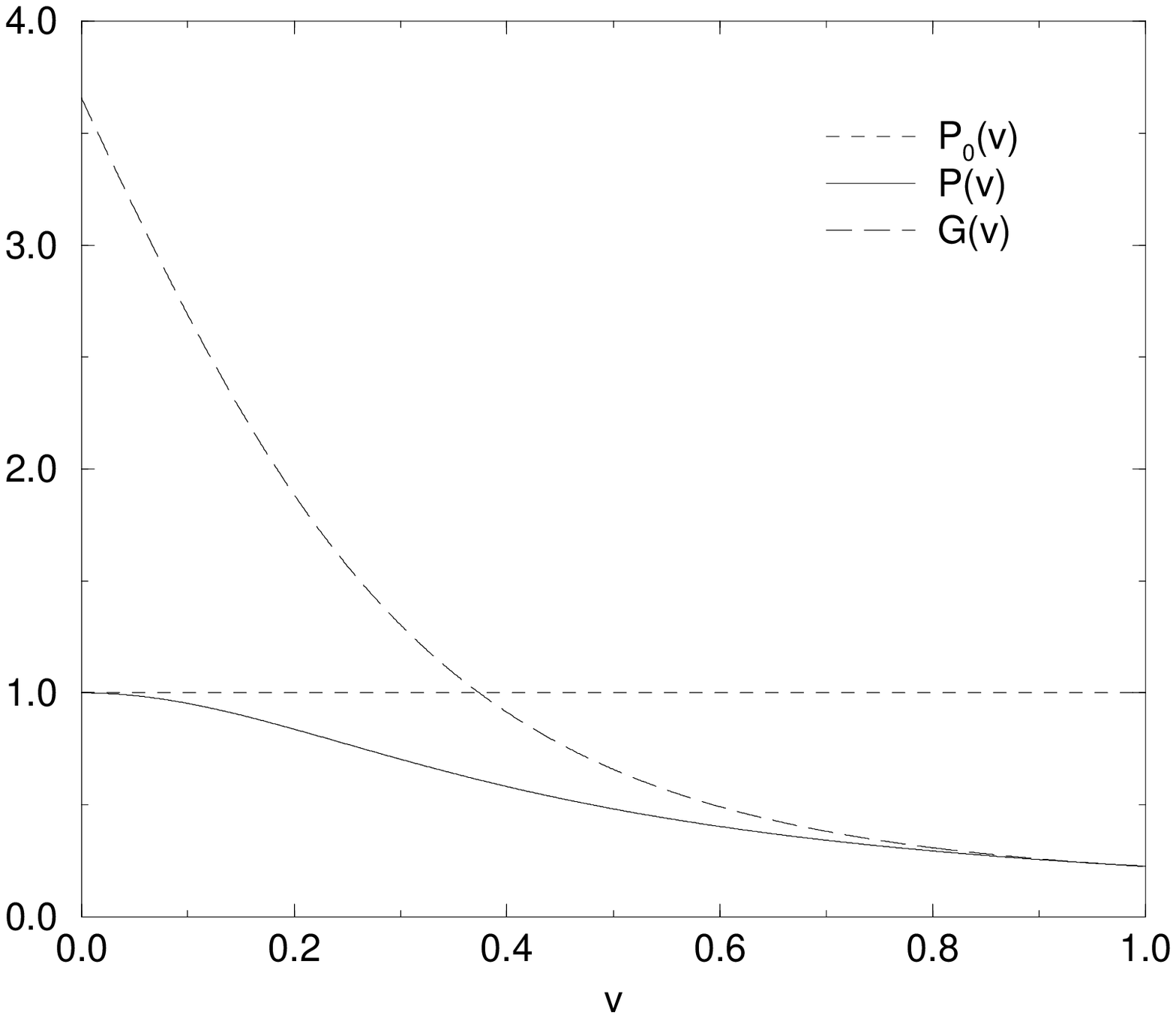}}
\noindent{\small {\bf Fig.2 }
  Velocity distributions in the case of a uniform initial distribution
  $P_0(v)=1$, for $R=10$.}
\end{figure}

\subsection{Uniform Cluster Distribution}

Consider the uniform final cluster distribution $P(v)=c$.  This
inverse problem is simple as all quantities can be obtained
explicitly.  From Eq.~(\ref{Q}), the auxiliary function is
$Q(v)=R^{-1}+{1\over 2}cv^2$ and from Eq.~(\ref{Qrho}) the initial
distribution reads

\begin{equation}
P_0(v)=c\left[1+{1\over 2} Rcv^2\right].
\end{equation}
The overall initial concentration is unity, thereby relating 
$R$ and $c$ via $1=c+{1\over 6}Rc^2$. The flux is calculated from 
Eq.~(\ref{j}), 
\begin{equation}
\label{Jlambda}
J={(3+\lambda)\sqrt{\lambda}\tan^{-1}\sqrt{\lambda}+
\lambda-\ln(1+\lambda)\over 3R},
\end{equation}
with $\lambda={1\over 2}Rc=(3/2)\left[\sqrt{1+2R/3}-1\right]$.  These
explicit solutions agree with our low and high $R$ predictions.  For
instance, when $R\gg 1$ we find $\langle m\rangle \sim R^{1/2}$ and
$J\sim R^{-1/4}$.  If we look at the initial distribution,
$P_0(v)\cong (6/R)^{1/2}+3v^2$, then the constant part is negligible
and the distribution corresponds to the $\mu=2$ case of the power-law
distribution (\ref{mu}).  For this case the size exponent is
$\alpha=1/2$ and the flux exponent is $\gamma=1/4$, see (\ref{mav})
and (\ref{j2}), in agreement with our findings.

Substituting $P_0(v)$ and $Q(v)$ in Eq.~(\ref{pvv}) and performing the
integration gives the joint distribution

\begin{eqnarray}
\label{Plambda}
P(v,v')&=&
2R^{-1}\lambda^2{(v-v')(1-\lambda v v')\over 1+\lambda v'^2}\\
+2R^{-1}\lambda^{3/2}(1&+&\lambda v^2)
\left[\tan^{-1}(\lambda^{1/2}v)-\tan^{-1}(\lambda^{1/2}v')\right].\nonumber
\end{eqnarray}
A direct integration of the joint distribution confirms the
conservation law (\ref{norm}), thus providing a useful check of
self-consistency.  The joint velocity distribution is linear in the
velocity difference for small $v$ and $v'$. This is reminiscent of the
small collision number behavior of Eq.~(\ref{pvv1}).  As the
velocity difference increases, significant curvature develops (see
Fig.~3). 

\begin{figure}
\centerline{\epsfxsize=8.5cm \epsfbox{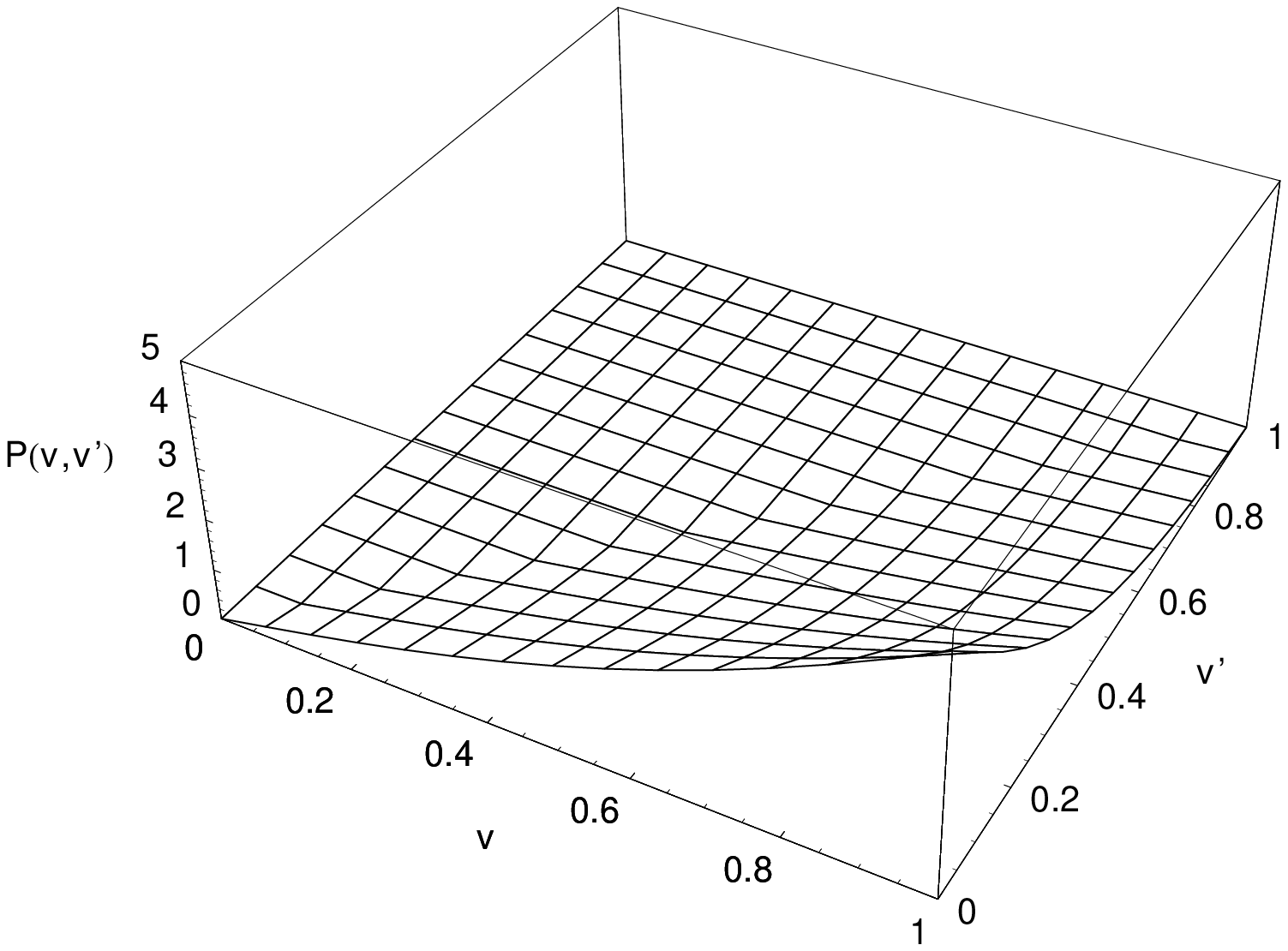}}
\noindent{\small {\bf Fig.3 }
The joint velocity distribution for the uniform final distribution 
case with $R=10$.}
\end{figure}

\subsection{Discrete Velocity Distribution}

The results formulated for continuous distributions can be used to
study the special case of discrete velocity distribution as
well.  Here we quote the results in terms of the original
(non-dimensionless) quantities. Consider the intrinsic velocity
distribution

\begin{equation}
\label{rho-dis}
P_0(v)=\sum_{i=1}^n c_i\delta(v-v_i),
\end{equation}
with $v_1<v_2<\cdots<v_n$.  We denote by $p_i$ the discrete
counterpart of the cluster velocity distribution, e.g., $P(v)=\sum_{i=1}^n
p_i\delta(v-v_i)$.  The steady state condition of Eq.~(\ref{pv}) reads
\begin{equation}
\label{pi}
p_i\left[1+t_0\sum_{j=1}^{i-1}(v_i-v_j)p_j\right]=c_i.
\end{equation}
Substituting the intrinsic velocity distribution and solving
iteratively, we get 
\begin{eqnarray}
\label{p123}
p_1&=&c_1\nonumber\\
p_2&=&{c_2\over 1+c_1(v_2-v_1)t_0}\\
p_3&=&{c_3\over 1+c_1(v_3-v_1)t_0+{c_2(v_3-v_2)t_0\over
    1+c_1(v_2-v_1)t_0}}\nonumber
\end{eqnarray}
etc. Rather than a solution to a differential equation, the steady
state solution is in the form of an explicit continued fraction. This
expression involves the initial distribution and the velocity
differences, and can be useful to analyze data in a histogram form. In
a similar way, explicit expressions can be obtained for the rest of
the steady state properties.

\section{Discussion}

An important property, the cluster size distribution is absent from
our treatment so far \cite{Gavrilov}. Naturally, the size and the
velocity of a cluster are strongly correlated and one must consider
$P_m(v)$, the distribution of clusters of size $m$ and velocity $v$.
The joint cluster size-velocity distribution obeys the master equation

\begin{eqnarray}
\label{pmvt}
{\partial P_m(v)\over\partial t}
&=&R^{-1}[mP_{m+1}(v)-(m-1)P_m(v)]\nonumber\\
&+&R^{-1}\delta_{m,1}[P_0(v)-P(v)]-F(v)P_m(v)\\
&+&\int_v^{\infty} dv' (v'-v)
\sum_{j=1}^{m} P_j(v')P_{m-j}(v)\nonumber
\end{eqnarray}
which applies for all $m\geq 1$.  Terms proportional to $R^{-1}$
account for escape, while the rest represent collisions.  The factor
$F(v)=\int_0^{\infty} dv'|v-v'| P(v')$ measures the overall collision
rate experienced by a $v$-cluster, and is reminiscent of kinetic
theory.  Summing Eqs.~(\ref{pmvt}), one recovers the rate equation
(\ref{pvt}) for $P(v)=\sum_m P_m(v)$.  On the other hand, integration
over the entire velocity range does not reduce Eqs.~(\ref{pmvt}) to a
closed system of rate equations for the cluster size distribution
$P_m=\int dv P_m(v)$. Therefore, the entire joint distribution is
needed to determine $P_m$.  Additionally, we note that in
Eqs.~(\ref{pvt}) and (\ref{pvvt}), the integration limits include only
slower velocities, a feature that considerably simplifies the analysis.
This property is lost for Eqs.~(\ref{pmvt}), thereby putting
analytical solution out of reach.

Nevertheless, a leading order analysis is still possible for low collision
numbers. In the limit $R\ll 1$, we find 
\begin{eqnarray}
P_1(v)&\cong& P_0(v)\left[1-R\int_0^{\infty}dv'\,
  |v-v'|P_0(v')\right],
\nonumber \\
P_2(v)&\cong& RP_0(v)\int_v^\infty dv'\,(v'-v)P_0(v'),\\
P_m(v)&\cong& R^{m-1}\tilde{P}_m(v).\nonumber
\end{eqnarray}
Heuristically, clusters with $m$ cars are created by $m-1$ collisions
and a factor $R$ is generated in each collision. Although the
functions $\tilde{P}_m(v)$ are quite complicated, the overall
prefactor $R^{m-1}$ suggests an exponential cluster size distribution
in the dilute limit.

In the special case of a bimodal velocity distribution, a solution is
possible. The structure of clusters here is simple: A cluster of size
$m$ consists of a leading slow car and $m-1$ fast cars behind it.  The
rate equation (\ref{pmvt}) simplifies considerably, and a Poisson size
distribution is found $P_m\propto e^{-f}f^{m-1}/(m-1)!$.  The
collision rate $f$ is equal to the product of the escape time, the
velocity difference, and the fast car concentration. This steady state
distribution satisfies a detailed balance condition as the escape rate
and the collision rate are equal microscopically, $(m-1)P_m=fP_{m-1}$.
Thus, an equilibrium steady state is reached.  However, in general, a
nonequilibrium steady state is approached with the collision rate and
the escape rate balancing only macroscopically. This is seen by noting
that the cluster size may increase by an arbitrary number due to
collisions, but can decrease only by one due to escape.
  
Further investigation of the collision term in the rate equation
will be useful as well. In the no escape case $R^{-1}=0$, the exact
Boltzmann equation
\begin{equation}
{\partial P(v,t)\over\partial t}=-P(v,t)\int_0^v dv' (v-v')P_0(v') 
\end{equation}
is different from our master equation as $P_0(v')$ replaces $P(v',t)$
in the integrand\cite{eps}. This seemingly small difference is
 important as it shows that the system remembers the initial
state. We argue that escape, no matter how small, induces mixing and
acts to erase this memory, and therefore Eq.~(\ref{pvt}).  It still
remain, however, to establish quantitatively how appropriate is this
mean field assumption. 

The model and the results presented above can be generalized to study
other traffic situations.  First, a multilane flow can be treated as a
system of coupled one lane flows. Escape naturally couples neighboring
lanes.  Second, a natural generalization is to heterogeneous
situations where passing is allowed only in a fraction $r$ of the
road.  We expect that for regular distribution of these passing
segments the problem should reduce to the homogeneous case with a
renormalized collision number $R/r$. The most challenging question 
appears to be the role played by the escape mechanism.  We
consideredthe case where all cars are equally likely to escape.  This
assumption simplified the master equation considerably as the escape
term is linear in $P(v)$, and thus, is exact. The complementary case
where only the first car in the cluster can escape is interesting as
well. For low collision numbers, large clusters are unlikely, and the
behavior is independent of the escape mechanism. However, for high
collision numbers the escape mechanism becomes weaker and larger
clusters form.  Indeed, a scaling argument along the lines of
Eq.~(\ref{m}) gives $\langle m \rangle \sim R$ in the car-cluster
dominated regime.

In conclusion, despite the simplifying assumptions made, the suggested
model results in quite realistic behavior. The overall picture is both
familiar and intuitive: due to the presence of slower cars, clusters
form and the overall flux is reduced.  For heavy traffic, the
characteristics of the flow are solely determined by the distribution
of slow cars. A single dimensionless parameter, the collision number
$R$ ultimately determines the nature of the steady state. The
stationary distributions obtained analytically provide a simple
practical recipe for calculating the flow properties for arbitrary
intrinsic distributions. It will be interesting to analyze observed
traffic data using these theoretical tools.  

\vspace{.1in}

We thank I.~Daruka and S.~Redner for useful discussions. PLK thanks
the CNLS for its hospitality and the ARO for financial support.

\end{multicols}

\end{document}